# Can participation in a hackathon impact the motivation of software engineering students? A preliminary case study analysis

Allysson Allex Araújo
Federal University of Cariri
Juazeiro do Norte, CE, Brazil
allysson.araujo@ufca.edu.br

Marcos Kalinowski
Pontifical Catholic University of Rio de Janeiro
Rio de Janeiro, RJ, Brazil
kalinowski@inf.puc-rio.br

Maria Teresa Baldassarre
University of Bari
Bari, BA, Italy
mariateresa.baldassarre@uniba.it

## ABSTRACT

[Background] Hackathons are increasingly gaining prominence in Software Engineering (SE) education, lauded for their ability to elevate students' skill sets. [Objective] This paper investigates whether hackathons can impact the motivation of SE students. [Method] We conducted an evaluative case study assessing students' motivations before and after a hackathon, combining quantitative analysis using the Academic Motivation Scale (AMS) and qualitative coding of open-ended responses. [Results] Pre-hackathon findings reveal a diverse range of motivations with an overall acceptance, while post-hackathon responses highlight no statistically significant shift in participants' perceptions. Qualitative findings uncovered themes related to networking, team dynamics, and skill development. From a practical perspective, our findings highlight the potential of hackathons to impact participants' motivation. [Conclusion] While our study enhances the comprehension of hackathons as a motivational tool, it also underscores the need for further exploration of psychometric dimensions in SE educational research.

## 1 INTRODUCTION

Becoming a proficient software engineer involves diverse educational pathways and training approaches [13]. From formal education programs to on-the-job training, aspiring software engineers have numerous avenues to acquire the requisite skill set. However, a reflection remains: How can we best motivate students in their journey to become competent software engineers?

In the field of Software Engineering (SE) education, it is essential to investigate strategies to enhance student motivation [20]. While technical knowledge is crucial, motivation also plays a significant role in sustaining long-term learning [12]. Therefore, there is a need to assess the effectiveness of innovative educational approaches, such as hackathons, in fostering motivation among SE students. In summary, a hackathon is an intensive collaborative event where participants, typically software developers, engage in focused, time-limited programming and problem-solving activities [17, 18]. Hackathons usually range from a few hours to a few days and are often organized around a challenge where the participants can work in teams to develop and present software solutions [19]. Hackathons could play a vital role in SE education by bridging the gap between theoretical concepts and practical application, promoting skills development, and preparing students for the dynamic and collaborative nature of the software industry [1, 8, 14]. How- ever, while hackathons have gained considerable popularity as a collaborative and immersive learning experience, their role as a motivational tool in SE education remains underexplored. From this need, our study seeks to address the following research question: "Can participation in a hackathon impact the motivation of software engineering students?". To explore this research question, we have designed a case study conducted at a federal university in Brazil to investigate the role of a hackathon in motivating SE students.

The hackathon was conducted in a hybrid format, combining both online and in-person activities, and spanned over seven days. Unlike the majority of studies focusing on hackathons as a pedagogical strategy for SE education [1, 8, 11, 17–19], our research adapted a widely acknowledged psychometric instrument, called Academic Motivation Scale (AMS) [22], to assess student motivation. Using validated psychometric instruments in SE research has been advocated [5], and recent critical reviews have indicated the importance of such instruments for maturing empirical SE investigations [6, 10]. By utilizing this psychometric evaluation, we aim to uncover insights into effective strategies for enhancing motivation and engaging SE students in their educational journey.

Hence, the contribution of this study is twofold. For academia, by providing an exploration of SE students' motivations within the context of innovative educational approaches such as a hackathon, using the AMS for investigating motivation dynamics, yielding insights that inform curriculum design and educational strategies. For practice, our findings highlight the potential of hackathons to impact participants' perceptions and motivations, suggesting the importance of incorporating proper team dynamics and focusing on skill development. These implications can help event organizers and educators create more engaging and impactful hackathon experiences that align with the demands of the SE industry.

The paper is structured as follows: Section 2 reviews related work; Section 3 details the case study design, followed by a presentation of preliminary results (Section 4) and discussion (Section 5). The paper concludes in Section 6 with final remarks.

## 2 RELATED WORK

Hackathons have been attracting considerable attention from SE researchers. For example, Porras et al. [14] analyzed a decade of hackathons and codecamps, elucidating the practical differences and benefits of each approach. In turn, Steglich et al. [19] focused on understanding the motivations, perceptions, and SE practices adopted by students participating in hackathons. They provided observations on the factors influencing student engagement and collaboration. More recently, Steglich et al. [18] delved into the development of professional skills and collaboration among students in an online hackathon, shedding light on the unique challenges and opportunities presented by virtual environments.'

Sadovykh et al. [17] emphasized the importance of integrating hackathons to bridge the gap between academic programs and industry needs. They also highlighted the benefits of aligning educational activities with industry practices. Complementing these findings, Gama et al. [8] presented a method for incorporating hackathons into undergraduate courses, focusing on the ideation process and its impact on learning. Kumalakov et al. [11] also proposed using hackathons as a teaching tool in SE courses, particularly in introductory classes, showcasing the potential to enhance student learning experiences. Lastly, Afshar et al. [1] examined the impact of hackathons on students' learning outcomes and code quality, providing discussions into the educational benefits.

Building upon the existing literature, our work aims to investigate the role of a hackathon as a motivational tool for SE students, specifically focusing on the assessment of motivation using a validated psychometric instrument, the Academic Motivation Scale (AMS). The AMS is the English translation of the *Echelle de Motivation en Education* [22]. Based on self-determination theory, this 28-item instrument is divided into seven subscales, reflecting one subscale of amotivation, three ordered subscales of extrinsic motivation (external, introjected, and identified regulation), and three distinct, unordered subscales of intrinsic motivation (intrinsic motivation to know, to accomplish things, and to experience stimulation) [4]. In summary, *intrinsic motivation* is the drive to pursue an activity simply for the pleasure or satisfaction derived from it, *extrinsic motivation* refers to pursuing an activity out of a sense of obligation, or as a means to an end, and, finally, *amotivation* is the absence of intent or drive to pursue an activity [21].

This study stands apart from previous SE research on hackathons, which mainly examined pedagogical methods, online formats, and skill outcomes. Our research distinguishes itself by centering on motivational aspects and being the first to explore the idea of approaching the AMS for a hackathon empirical evaluation. This validated instrument enables us to discuss how hackathons influence motivation among SE students. Hence, this short paper tackles the use of a hackathon in SE education along with an interdisciplinary approach supported by preliminary results that cover thought-provoking discussions on motivational aspects.

## 3 CASE STUDY DESIGN

This investigation concerns an evaluative case study with students as the unit of analysis for assessing their motivation. We followed the methodological guidelines provided by Runeson and Höst [16]. These guidelines emphasize **five main aspects** addressed below.

This study aims to address the following **research question** (1): *Can participation in a hackathon impact the motivation of software engineering students?* This research question guides our investigation, delving into the intricate relationship between hackathons and student motivation. By exploring this dynamic, we aim to contribute with a better understanding of the existing body of knowledge on effective SE education and training strategies.

The **case selection** (2) involved a hackathon at a federal university in Brazil's northeast, which aimed to bring together students to address a real-world problem. The purposeful selection of this case was based on its relevance to the research aim of investigating the impact of a hackathon as a motivational tool for SE students. The participants were selected from the pool of bachelor students who expressed interest in participating in the hackathon. The selection criteria considered their enrollment in the event.

The **data collection** (3) process encompassed two major phases: pre-hackathon and post-hackathon. Prior to the beginning of the event, participants were required to complete a questionnaire. This questionnaire sought demographic information, including age, gender, course of study, duration of enrollment at university, self-assessed software development proficiency (rated on a Likert scale of 1 to 5), self-assessment of teamwork abilities (rated on a Likert scale of 1 to 5), and an open-ended question to gauge any prior experience with hackathons. Additionally, participants were administered the AMS (rated on a Likert scale of 1 to 7) as part of the pre-hackathon data collection. Considering that we dealt with Brazilian students, we administered the validated translation into Brazilian Portuguese of the AMS [3]. As depicted in Table 1, the questions were also slightly adapted to the context of a hackathon. Subsequent to the conclusion of the hackathon, participants were presented with a follow-up questionnaire, which also featured a series of scaled questions requiring self-assessment. The questions were designed to gauge the extent to which participation in the hackathon contributed to academic engagement, stimulated creativity and innovation, fostered a collaborative and teamwork-oriented environment, and facilitated the practical application of knowledge gained in the classroom. The scale for each question ranged from 1 to 7. An open-ended question was included to elicit additional comments based on the participants' experiences during the event. Again, participants were requested to complete the AMS as part of the post-hackathon data collection process.

The **data analysis** (4) encompassed both quantitative and qualitative approaches. Quantitative data derived from the pre- and post-hackathon questionnaires, encompassing demographic information and AMS scores, underwent descriptive statistics analysis. Measures such as means, standard deviations, and frequency distributions were employed. Specifically, inferential statistics, including non-parametric tests like the Mann-Whitney U Test and Cohen's d statistics, were also utilized to examine the AMS scores and assess any significant changes before and after the hackathon. In contrast, the qualitative data obtained from open-ended responses underwent open coding as a qualitative data analysis technique.

Different procedures were implemented to ensure the **validity of the findings** (5). Hence, using a validated psychometric instrument enhanced the measurement validity of the motivation constructs. Secondly, data triangulation was employed by collecting data questionnaires, open-ended questions, and a literature review. This approach allowed for cross-validation of the findings and increased the reliability of the results [15]. To uphold the standards of methodological rigor, reliability, and validity, we adhered to the case study protocol as recommended by Runeson and Höst [16]. All the data underpinning this study is openly available via our supporting repository [2], ensuring transparency and accessibility.

## 4  PRELIMINARY RESULTS

Regarding the case study description, the hackathon lasted seven days, from May 18 to May 24, 2023, and involved ten teams of four students. The selection of teams was carried out through an application process based on the order of registration. The announcement of the selected teams was made on May 15, 2023. The hybrid nature of the week allowed for both online and in-person activities, including intense use of self-management skills. Participants had access to the hackathon's Discord server, where communication and collaboration occurred among team members, mentors, and event organizers. Depending on the team's specific needs, they were granted access to the university's laboratory facilities. Mentors, alumni of the university (and are currently working in the software industry), were assigned to each team to provide guidance and support in refining their projects.

The problem addressed during the hackathon was the development of a meeting tracking software for undergraduate thesis supervision. The hackathon's organizers identified this problem as an existing demand within the university. In summary, the software aimed to manage and record meetings (tasks to be done, logs, etc.) between supervisors and students while allowing course coordination to track their activities. The product requirements were communicated to the participants through a software requirements document. On the final day of the hackathon, each team was required to present a minimally functional prototype with front-end and back-end, following a slide deck that covered specific points in the format of a pitch presentation. Only one team opted not to present their pitch for personal reasons. The solutions were evaluated by a panel of five invited judges, who assessed criteria such as creativity and innovation, quality of design and usability, functionality and feasibility, and presentation (oral and visual). The team with the highest average score was chosen as the winner. Each winning team member received a personalized small trophy (3D printed in the university) and a sponsored gift voucher as a symbolic prize. All teams that presented their projects received a hackathon completion certificate.

From 40 enrolled participants, 34 responses (85%) were obtained for the pre-hackathon questionnaire, while 35 responses (87.5%) were collected post-hackathon. The gender distribution among the participants indicated that 88.23% were male and 11.76% were female. In terms of age, most respondents fell within the 18-21 age range, accounting for 79.40% of the participants. The 22-25 age range accounted for 14.70% of the participants, while the 26-30 and over 30 age ranges represented 2.90% each. Regarding the distribution of courses, 50% of the participants were from the Information Systems course, 47.10% were from Computer Science, and the remaining 2.90% were from Mining Engineering. Regarding enrollment duration at university, 26.50% had been enrolled for less than one year, 61.80% had been enrolled for one to three years, and the remaining 11.80% had been enrolled for over three years.

Table 1 summarizes the quantitative results. The columns 'Likert pre-hackathon' and 'Likert post-hackathon' feature a grouped distribution analysis to enhance data visualization. Instead of individual colors for each Likert scale, we have employed the following color scheme: scales 1, 2, and 3 are represented in orange, scale 4 in grey, and scales 5, 6, and 7 in blue. This table also presents the pre- and post-hackathon mean and standard deviation results, alongside indications of shifts in these values (↑ denoting an increase and ↓ a decrease). Additionally, we report the difference between post- and pre-hackathon mean values, p-value, and effect size.

From the data collected pre-hackathon, the statement "Honestly, I do not know why I participate in the hackathon" garnered significant attention, with 79.41% of respondents indicating a low correspondence (scale 1) to demotivation. Thus, a substantial proportion of students clearly understand their motivations. Conversely, the statement "I participate in the hackathon to challenge myself and show my ability to create innovative solutions" revealed a contrasting trend. Here, most of participants (over 80%) expressed a high correspondence (scales 6 and 7), indicating their extrinsic motivation to challenge themselves and exhibit innovative problem-solving skills. This finding underscores the hackathon's role in fostering an environment that stimulates personal growth.

The variance in participants' responses to different motivational statements evidences the multifaceted nature of their extrinsic motivation. For instance, the statement "I participate because it is what is expected of me as a student" had an average of 2.76 and a relatively high standard deviation of 1.88. This higher deviation suggests a divergence of opinions, potentially indicating varying perceptions of societal expectations and peer influence. The statement "I participate in the hackathon so as not to miss out on the learning and networking opportunities it offers" resonated strongly, with a high average of 6.23 and a relatively low standard deviation of 1.07. This result suggests a collective recognition of the hackathon's potential for skill enhancement and networking.

Concerning the survey data obtained post-hackathon, we found no statistically significant changes in participants' motivations, as indicated by the p-values and effect sizes. This finding could be attributed to factors like the relatively small sample size, the complexity of measuring motivation, and the possibility that participants' pre-hackathon motivations for attending were already high (they voluntarily registered to participate). Moreover, this outcome unravels the challenging nature of assessing motivation shifts as well as the potential need for a more sensitive psychometric instrument specifically tailored to the context of hackathons in SE education [10]. Hence, while the AMS provided valuable findings, its generic nature might not fully capture the intricacies of motivations in this unique setting.

| Adapted AMS questions | Likert pre-hackathon | Mean ± std. dev. pre-hackathon | Likert post-hackathon | Mean ± std. dev. post-hackathon | Change | Mean difference | p-value | Effect Size |
|---|---|---|---|---|---|---|---|---|
| Honestly, I do not know why I participate in the hackathon | | 1.35 ± 0.80 | | 1.34 ± 0.82 | ↓ | -0.01 | 0.849 | 0.012 |
| I participate in the hackathon because it's a mandatory activity | | 1.08 ± 0.28 | | 1.11 ± 0.39 | ↑ | 0.03 | 0.992 | 0.087 |
| I participate in the hackathon to avoid losing points or being penalized | | 1.14 ± 0.54 | | 1.11 ± 0.39 | ↓ | -0.03 | 0.984 | 0.064 |
| For the pleasure I feel when getting involved in projects and collaborating with other students during the hackathon | | 6.26 ± 0.94 | | 6.11 ± 0.95 | ↓ | -0.15 | 0.689 | 0.144 |
| I participate in the hackathon to challenge myself and showcase my ability to create innovative solutions | | 6.23 ± 1.26 | | 6.26 ± 1.40 | ↑ | 0.03 | 0.667 | 0.020 |
| I participate in the hackathon so I do not miss out on the learning and networking opportunities it offers | | 6.23 ± 1.05 | | 6.06 ± 1.33 | ↓ | -0.17 | 0.689 | 0.142 |
| I feel like I am wasting my time in the hackathon | | 1.23 ± 0.87 | | 1.43 ± 0.77 | ↑ | 0.20 | 0.226 | 0.239 |
| I participate because it's what's expected of me as a student | | 2.76 ± 1.84 | | 2.63 ± 1.74 | ↓ | -0.13 | 0.889 | 0.073 |
| I used to have good reasons to participate in hackathons, but now I have doubts about whether I should continue | | 1.35 ± 0.87 | | 2.06 ± 1.87 | ↑ | 0.71 | 0.238 | 0.480 |
| To prove to myself that I have the skills and intelligence to face challenges in the technology field | | 5.80 ± 1.51 | | 5.60 ± 1.62 | ↓ | -0.20 | 0.390 | 0.125 |
| I participate in the hackathon because attendance is mandatory | | 1.08 ± 0.28 | | 1.11 ± 0.32 | ↑ | 0.03 | 0.857 | 0.114 |
| Because participating in academic activities is a privilege and an opportunity for growth | | 5.85 ± 1.49 | | 6.17 ± 1.32 | ↑ | 0.32 | 0.390 | 0.210 |
| I do not understand why I should participate in hackathons | | 1.38 ± 1.00 | | 1.23 ± 0.86 | ↓ | -0.15 | 0.516 | 0.162 |
| I participate in the hackathon to obtain a certificate of participation | | 3.08 ± 1.8 | | 2.80 ± 2.09 | ↓ | -0.28 | 0.424 | 0.094 |
| Participating in the hackathon makes me feel important when I succeed in projects | | 5.11 ± 1.67 | | 4.91 ± 2.36 | ↓ | -0.20 | 0.741 | 0.094 |
| I do not know, and I do not understand the purpose of participating in the hackathon | | 1.26 ± 0.60 | | 1.20 ± 1.01 | ↓ | -0.06 | 0.308 | 0.071 |
| Because, for me, the hackathon is a enjoyable and stimulating experience | | 5.85 ± 1.37 | | 5.91 ± 1.61 | ↑ | 0.06 | 0.459 | 0.043 |
| Because the hackathon is an opportunity to access knowledge and new technologies | | 6.14 ± 1.03 | | 6.03 ± 1.50 | ↓ | -0.11 | 0.674 | 0.085 |
| I do not realize what difference participating in the hackathon makes | | 1.38 ± 1.00 | | 1.14 ± 0.54 | ↓ | -0.24 | 0.395 | 0.289 |
| Because I want to prove to myself that I can succeed in developing solutions. | | 5.67 ± 1.67 | | 5.83 ± 1.70 | ↑ | 0.16 | 0.603 | 0.093 |
| Because I have a strong interest and pleasure in participating in the hackathon | | 6.00 ± 1.26 | | 5.77 ± 1.59 | ↓ | -0.23 | 0.818 | 0.157 |
| Because I consider that keeping a record of participation is important to track my progress and learning during the hackathon. | | 5.17 ± 1.99 | | 5.11 ± 2.03 | ↓ | -0.06 | 0.936 | 0.027 |
| I want to avoid being seen as an uninterested or disengaged participant in the hackathon | | 4.00 ± 2.33 | | 3.91 ± 2.47 | ↓ | -0.09 | 0.865 | 0.037 |
| I participate in the hackathon because attending the activities is mandatory | | 1.15 ± 0.49 | | 1.14 ± 0.42 | ↓ | -0.01 | 0.881 | 0.021 |
| If participation wer not mandatory, few students would participate in the hackathon | | 1.58 ± 1.35 | | 1.71 ± 1.65 | ↑ | 0.13 | 0.881 | 0.085 |
| Because studying and participating in projects like the hackathon broaden my academic and professional horizons | | 6.29 ± 1.17 | | 6.00 ± 1.66 | ↓ | -0.29 | 0.726 | 0.247 |
| I participate in the hackathon because it was a personal choice that aligns with my interests and goals | | 6.23 ± 1.05 | | 6.06 ± 1.58 | ↓ | -0.17 | 0.928 | 0.124 |
| I participate in the hackathon because while I am participating, I do not have to work | | 1.14 ± 0.42 | | 1.37 ± 0.96 | ↑ | 0.23 | 0.549 | 0.306 |
| My friends are the main reason I participate in the hackathon | | 3.85 ± 0.78 | | 4.34 ± 2.11 | ↑ | 0.49 | 0.194 | 0.247 |

**Table 1: Overview of the quantitative results.**

The statement "Honestly, I do not know why I participate in the hackathon" exhibited a response shift after the hackathon. Pre-hackathon, a majority of students (79.41%) selected scale 1, suggesting confidence about their motives. After the hackathon, this percentage increased to 82.85%, which signifies a subtle shift in the understanding of their engagement. Another noteworthy shift was observed in the statement "My friends are the main reason I participate in the hackathon". Post-hackathon was a significant increase in the average score, rising from 3.85 to 4.34, which evidence the extrinsic motivation by peer relationships as introjected regulation.

Another statement that exhibited a change post-hackathon was "I participate in the hackathon to challenge myself and show my ability to create innovative solutions". The responses showed that a significant number of students (68.57%) strongly agreed with this motivation, representing an increase from the pre-hackathon responses (almost 10%) regarding scale 7. The statement "For the pleasure I feel when getting involved in projects and collaborating with other students during the hackathon" exhibited a considerably high average value (6.26) post-hackathon. This result indicates an alignment with intrinsic motivations. Another intriguing shift was seen in the response to "I once had good reasons to participate in hackathons, but now I have doubts about whether I should continue participating". The average increased from 1.35 to 2.06, showcasing a shift in participants' demotivation regarding their initial doubts about continuing their participation. Notably, the standard deviation increased from 0.88 to 1.89, indicating a broader dispersion of responses following the hackathon experience.

As we can see, examining AMS results uncovers subtle shifts in student motivations toward hackathon participation. As demonstrated by the non-significant p-values and *Avg. difference* results, most motivations remained relatively stable, such as the pleasure of engaging in projects, skill showcase, and the expectation as a student. Others revealed some alterations that reflect a more profound engagement with the event. In particular, motivations related to friendship and educational growth exhibited more remarkable changes. Moreover, it is noteworthy that even when exclusively examining post-hackathon responses in isolation, without comparative analysis with pre-hackathon data, the results remain notably positive regarding students' motivation.

To complement this AMS analysis, we also addressed in our study six other supplementary Likert-based inquiries aimed at capturing other analytical dimensions, including students' self-awareness regarding teamwork and their software development proficiency, among others. Due to space constraints, we emphasize a relevant post-hackathon query herein: "On a scale of 1 to 7, to what extent did participation in the hackathon contribute to your academic engagement?". Remarkably, the responses yielded an average rating of 5.97, with 94.12% of the student cohort indicating a response falling within the range of 5 to 7. This result highlights that a vast part of students perceived a positive contribution to their academic engagement from participating in the hackathon.

In addition to the quantitative results gained from the Likert scale responses to the AMS, participants were allowed to provide open-ended comments based on their hackathon's experience. Qualitative open coding revealed four distinct codes (summarized in Figure 1), each shedding light on the participants' perspectives.

**Enriching Experience and Networking** highlights the event's success, surpassing participants' expectations and fostering collaboration. Participants lauded the hackathon for facilitating the exchange of experiences and knowledge, with a significant desire to participate again. This code emphasizes the event's role in cultivating a vibrant learning and networking environment. Moreover, **Time Constraints, Scheduling, and Complexity** refers to the event logistics. Participants voiced the need for well-timed scheduling, even though this constraint is a usual feature in hackathons. The challenge of managing project complexity within limited timeframes also emerged, signaling the importance of project planning.

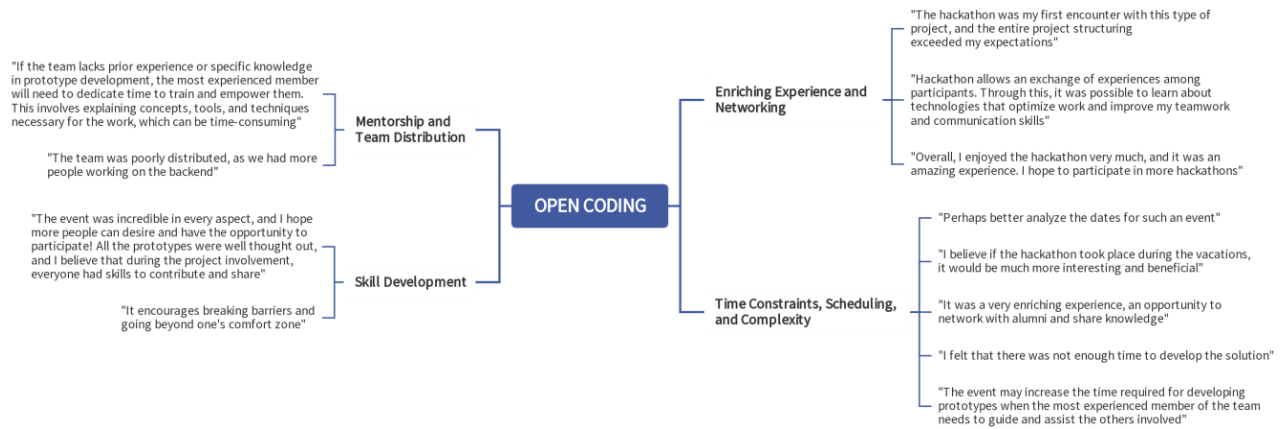

**Figure 1: Open coding covering the open-ended question made post-hackathon.**

**Team Dynamics** delves into team distribution and previous knowledge. Participants recognized that teams with varied skill levels necessitated additional training and mentoring, indicating a potential trade-off between skill diversity and project speed. Concerns regarding the uneven distribution of skills within teams were evident, emphasizing the need for balanced team compositions. Lastly, **Skill Development** encapsulates the transformative nature of the hackathon. Participants praised the hackathon for cultivating a culture of personal growth, breaking comfort barriers, and fostering skill diversification. This preliminary qualitative analysis encompasses the multifaceted impact of the hackathon.

## 5 DISCUSSION

Towards answering our research question, the quantitative analysis of students' motivations prior to the hackathon yielded a range of insights into the factors driving their motivation. It is worth highlighting the acceptance by students of hackathons as fostering personal growth, as well as the influence of friendship. In line with these findings, qualitative analysis unveiled an additional layer of discussion, exposing themes like proper product planning, diverse teams, and hackathons' capability to leverage skill development and a growth mindset. These results offer valuable implications for hackathon pedagogical design and emphasize its broader potential in increasing students' motivation and skills enhancement.

In summary, the observed scenario in post-hackathon perceptions emphasizes the capacity of such events to be appreciated by the students [1, 8, 11, 17–19]. Incorporating product management capabilities could enhance collaboration and knowledge sharing while focusing on skill development aligns with industry needs towards adaptable and innovative graduates [9]. Although this study provides preliminary results into participants' experiences, it also suggests the necessity for further exploring psychometric dimensions within SE education [10]. Lastly, integrating the AMS played a pivotal role, yet the absence of statistically significant differences denotes the call for replications. This research gap highlights the need for a deeper exploration to unravel the intricacies of SE students' motivations [7].

## 6 CONCLUSIONS AND FUTURE WORK

This study delves into the motivations of Software Engineering (SE) students participating in a hackathon. The research examined the perceptions of participants both before and after the event, providing an understanding of the factors driving their participation. The pre-hackathon phase reveals a diverse range of motivations, including intrinsic growth aspirations and extrinsic factors, reflecting the complex nature of participants' intent. Post-hackathon responses highlight a remaining positive outlook on their motivations. These findings align with existing evidence emphasizing the benefits of hackathons in promoting intrinsic motivation, collaborative learning, and skill development. In the context of SE education, hackathons are particularly valuable as they equip students with technical expertise and soft skills demanded by industry.

However, this preliminary research has limitations. The reliance on self-reported data and a single-institution focus may limit the broader applicability of findings. The lack of a control group also prevents definitive conclusions about the causality of hackathons on motivation. Moreover, the lack of statistical significance needs cautious interpretation. This issue suggests the potential influence of factors like sample size on the ability to detect significant changes. However, students reported a high academic engagement, with an average rating of 5.97 out of 7. Despite these limitations, the use of the Academic Motivation Scale (AMS) enabled us to enhance our comprehension of hackathons as a motivational tool, also furnishing actionable inputs to explore avenues for improving the design and impact of hackathons in SE education.

Our study adds to the existing groundwork on behavioral aspects in SE education and paves the road for future studies exploring motivational factors. As for future work, we may expand our scope by investigating a wider range of hackathons and diverse settings (different institutions, skill levels, etc.).